\documentclass[aps, prl, twocolumn,showpacs,preprintnumbers,amsmath,amssymb]{revtex4}

\usepackage{graphicx}
\usepackage{dcolumn}
\usepackage{bm}
\newtheorem{law}{Proposition}

\begin{document}

\title{Maximum Speedup in Quantum Search : $O(1)$ Running Time}

\author{Joonwoo Bae$^{1}$} 
\email{Joonwoo.Bae@upc.es}
\author{ Younghun Kwon$^{1,3}$}
\email{yyhkwon@hanyang.ac.kr}
\affiliation{$^1$ Department of Physics, Hanyang University, \\Ansan, Kyunggi-Do, 425-791, South Korea\\}

\affiliation{$^2$ Department of Physics, Unversity of Rochester, Bausch and Lomb Hall, P.O. Box 270171,600 Wilson Boulevard, Rochester, NY 14627-0171 \\}

\date{\today}

\begin{abstract} 
Recently the continuous time algorithm based on the generalized quantum search Hamiltonian  was presented. In this letter, we consider the running time of the generalized quantum search  Hamiltonian. We provide the surprising result that the maximum speedup of quantum search in  the generalized Hamiltonian is the $O(1)$ running time regardless of the number of total states. It seems to violate the proof of Zalka that the quadratic speedup is optimal in quantum search. However the argurment of Giovannetti et al. that a quantum speedup comes from the interaction between subsystems(or, equivalently entanglement) (and  is concerned with the Margolus and Levitin theorem) supports our result. 
\end{abstract}

\maketitle

Quantum computation is an epoch-making idea to change concept of computation. Since David  Deutsch showed that  characteristics of quantum physics such as superposition and entanglement can be used to process information more efficiently than any classical machine\cite{14} , there have been much progress in quantum computation. One of them is Grover's quantum search algorithm which takes $O(\sqrt{N})$ trials while a known classical search algorithm needs O(N).\cite{9}\cite{10}  Grover offered the quantum search algorithm with quantum gates. On the other hand  Farhi and Gutmann provided the following Hamiltonian performing quantum search  \cite{1} 

\begin{displaymath}
H=E( |w \rangle \langle w| +  |\psi \rangle \langle \psi|) \nonumber \\
\end{displaymath}

The Hamiltonian may correspond to Grover algorithm.\cite{13} Recently, the generalized quantum search Hamiltonian was proposed.\cite{3} 

\begin{eqnarray}
 H=E(|w \rangle \langle w| + |\psi \rangle \langle \psi|)  + \epsilon (e^{i\phi}|w \rangle \langle \psi| + e^{-i \phi} |\psi \rangle \langle w|) \nonumber \\
\end{eqnarray}

where, $E$ and $\epsilon$ are positive constants in unit of energy with the condition $E \geq \epsilon$ and $\phi$ is a constant phase.\cite{4} We denote the initial state as $|\psi \rangle = x|w\rangle +\sqrt{1-x^{2}}|r\rangle$, where $|w\rangle$ is the target state, $|r\rangle$ the orthogonal complement, and $x = \langle w|\psi\rangle(\approx 1/\sqrt{N})$. Let us consider the running time of the generalized quantum search Hamiltonian. Then we can find the surprising speedups involving the $O(1)$ time. One may think that the O(1) speedup violates the proof of Zalka that the quadratic speedup is optimal in quantum search. However the Margoulus and Levitin theorem and the recent argument of Giovannetti et al.\cite{6}\cite{7} help us  to understand the consistency of the maximum speedup.\cite{2} \\ 
 The running time of the algorithm based on the generalized quatnum search Hamiltonian is 

\begin{eqnarray}
T = \frac{\pi}{2}\frac{1}{[(Ex+\epsilon \cos\phi)^{2}+(1-x^{2})\epsilon^{2}\sin^{2}\phi]^{1/2}} \nonumber \\
\end{eqnarray}

The running time $T$ may be $O(\sqrt{N})$ for arbitrary $E$, $\epsilon$ and $\phi$ with $E > \epsilon$. We name the Hamiltonian having the above quadratic speedup as the type1 quantum search Hamiltonian.\\

$Type1.$ The quadratic speedup quantum search Hamiltonian \\

$H_{1} = E(|w \rangle \langle w|+|\psi \rangle  \langle \psi|)+\epsilon(e^{i\phi}|w \rangle \langle \psi|+e^{-i\phi}|\psi \rangle \langle w|)$ with $E>\epsilon$ \\

Next we consider the case of $E=\epsilon$. In this case, the running time is 

\begin{eqnarray}
T &=& \frac{\pi}{2} \frac{1}{E(xcos \phi +1)} \nonumber \\
  &=& \frac{\pi}{2E} (1+O(x\cos\phi)) \nonumber \\
\end{eqnarray}

 Choosing $\phi = \pm \frac{\pi}{2}$, we surprisingly have the $O(1)$ time as follows

\begin{displaymath}
T = \frac{\pi}{2E} = O(1)
\end{displaymath}

In this case, the running time has nothing to do with the total number of states $N$. This is the case of  the maximum speedup $O(1)$. We also name the Hamiltonian having the $O(1)$ speedup as the type2 quantum search Hamiltonian.\\

$Type2.$ The $O(1)$ time quantum search Hamiltonian \\

$H_{2} = E[|w \rangle \langle w|+|\psi \rangle \langle \psi| \pm i|w \rangle \langle \psi| \mp i|\psi \rangle \langle w|]$ \\

In addition, there are another two kinds of quantum search Hamiltonian, whose running time is reducing as the number of states becomes larger. We name the Hamiltonian as type3 quantum search Hamiltonian as follows: \\

$Type3-1.$ The first exponential speedup Hamiltonian \\

$H_{31} = E(|w \rangle \langle w|+|\psi \rangle \langle \psi|)+\epsilon(e^{i\phi}|w \rangle \langle \psi|+e^{-i\phi}|\psi \rangle \langle w|)$ where $\phi = \cos^{-1}(-Ex/\epsilon)$ , with the condition $E> \epsilon > Ex$\\

, whose running time is 

\begin{eqnarray}
T &=& \frac{\pi}{2}\frac{1}{[(1-x^{2})(\epsilon^{2}-(Ex)^{2})]^{1/2}} \nonumber \\ &=& O(\frac{1}{[(\epsilon^{2}-(Ex)^{2})]^{1/2}}) \nonumber \\
\end{eqnarray}

$Type3-2.$ The second exponential speedup Hamiltonian \\

$H_{32} = E(|w \rangle \langle w|+|\psi \rangle  \langle \psi| + e^{i\phi}|w \rangle \langle \psi|+e^{-i\phi}|\psi \rangle \langle w|)$ \\

, whose running time is 

\begin{eqnarray}
T &=& \frac{\pi}{2} \frac{1}{E(xcos \phi +1)}  \nonumber \\
  &=& \frac{\pi}{2E} (1+O(x\cos\phi))
\end{eqnarray}

$Type4.$ The exponential and quadratic speedup Hamiltonian \\

$H_{4} = E(|w \rangle \langle w|+|\psi \rangle  \langle\psi|) \pm \epsilon(i|w \rangle \langle \psi|-i|\psi \rangle \langle w|)$ \\

, whose running time is 

\begin{eqnarray}
T &=& \frac{\pi}{2} \frac{1}{\sqrt{(E^{2}-\epsilon^{2})x^{2}+\epsilon^{2}}} \nonumber \\
& = & O(max(\frac{1}{\epsilon}, \frac{1}{\sqrt{E^{2}-\epsilon^{2}}x}))  \nonumber \\
\end{eqnarray}

The running time $T$ is $O(1)$ if $Ex > \epsilon\sqrt{1+x^{2}}$, but $T = O(\sqrt{N})$ otherwise.\\
 The generalized quantum search Hamiltonian may find the target state ,with various speedups, involving the $O(1)$ time and the exponential.  The quantum search Hamiltonians can be classified to 4 types corresponding to their speedups. For instance, the Farhi and Gutmann's quantum search Hamiltonians is included in the type1. At the measurement time(or in the read-out time), the probability for Hamiltonians of type1, type2, and type 3 to obtain the target state may be $1-O(x^{2}) \approx 1-O(1/N)$, although their running times are different. The condition $\phi = n \pi$ is necessary for the probability one.\cite{4} Thus the perfect searching holds only for the Hamiltonians of type1 and type3. \\
 The $O(1)$ time in quantum search is indeed something exotic if one remember the Zalka's proof that the quadratic speedup of the Grover algorithm is optimal.\cite{12} Let us then consider the Margolus and Levitin theorem to resolve the maximum speedup.\cite{6} The theorem claims that, for a Hamiltonian whose lowest energy level is zero, it takes  at least a time $T_{\bot} \geq \pi/2E$ for a given state $|\psi\rangle$ to evolve an orthogonal state, where $E$ is the mean enegy, $E = \langle \psi|H|\psi \rangle$. Giovannetti et al. stated the following theorem concerning the Margolus and Levitin theorem\cite{7} \\

\begin{law}[Maximum Speedup]
Suppose that ,for a Hamiltonian $H$, a state of the Hamiltonian $|\eta \rangle$ is given. Then the minimum time for the state to evole to an orthogonal state is 

\begin{displaymath}
T_{\bot} \geq T(E,\triangle E) := max(\frac{\pi}{2E}, \frac{\pi}{2\triangle E})  \nonumber \\
\end{displaymath} 

where $E = \langle \eta |H| \eta \rangle$(mean energy) and $\triangle E = \sqrt{\langle \eta |(H-E)^{2}|\eta\rangle}$(standard deviation)
\end{law}

Applying the maximum speedup theorem to the generalized quantum search Hamiltonian, we obtain the minimum evolution time. The lowest energy level of the Hamiltonian is non-zero, so the time $T_{\bot}$ is 

\begin{eqnarray}
T_{\bot} & \geq & T(E,\triangle E)  \nonumber \\
& = & \frac{\pi}{2}\frac{1}{[(Ex+\epsilon \cos\phi)^{2}+(1-x^{2})\epsilon^{2}\sin^{2}\phi]^{1/2}}  \nonumber \\
\end{eqnarray} 

This minimum evolution time coincides with the running time of the algorithm based on the Hamiltonian. Thus, we have shown that the running time of the quatum search Hamiltonian is minimum. The works to minimize the lower bound of the evolution time $T_{\bot}$ are expected to be the same to the various speedups, the $O(1)$ time and the exponential. This implies that, although the $O(1)$ running time is somewhat surprising in view that Grover algorithm is optimal, but, by the theorem, expected in a continuous time algorithm based on Hamiltonian evolution. \\
 Let us revisit the Farhi and Gutmann Hamiltonian to consider the interaction between the target state and the initial one. The Farhi and Gutmann Hamiltonian can be written as 

\begin{eqnarray}
H & = & H_{w}+H_{r} + H_{int}  \nonumber \\
\end{eqnarray} 

where $H_{w} = E(1+x^{2})|w \rangle \langle w|$ and $H_{r} = E(1-x^{2})|r\rangle\langle r|$ are the free Hamiltonians and 

\begin{eqnarray}
H_{int} & = & E x\sqrt{1-x^{2}}(|w\rangle \langle r|+|r \rangle \langle w|) \nonumber \\
\end{eqnarray} 

is the interaction Hamiltonian. The  mean energy of this system is $E$ and the standard deviation of it $\triangle E = Ex$. Thus the minimum evolution time is $T_{\bot} = \pi/2Ex $, which coincides with the running time of Farhi and Gutmann Hamiltonian. This implies the running time of the Farhi and Gutmann Hamiltonian is the minimum. In the sense that $|w \rangle$ and $|r \rangle$ can be considered mutually negated states, the Hamiltonian can be explained by the entanglement dynamics argument of Giovannetti et al. This provides the point of view that quantum search is the process to evolve the initial state to the target state with the interaction between the two states. We then generalize the interaction between the two states, by  letting the following Hamiltonian  perform quantum search 

\begin{eqnarray}
H & = & H_{w}+H_{r} + H_{int} \nonumber \\
\end{eqnarray} 

\begin{eqnarray}
H_{w} & = & E_{1} |w\rangle \langle w|  \nonumber \\
H_{r} & = & E_{2} |r\rangle \langle r|  \nonumber \\
H_{int} & = & E_{3}(e^{i\varphi}|w\rangle \langle r|+e^{-i\varphi}|r \rangle \langle w|)  \nonumber \\
\end{eqnarray} 

where $H_{w}$ and $H_{r}$ are the free Hamiltonian and $H_{int}$ is the interaction Hamiltonian. We will exert to determine the unknowns $E_{1}$, $E_{2}$ and $E_{3}$ and the phase $\varphi$, in order that the Hamiltonian describes the general interaction between the initial state and the target state. An ideal quantum search algorithm should be written with the initial state $|\psi\rangle$($=x |w \rangle +\sqrt{1-x^{2}}|r\rangle$) and the target state $|w \rangle$. Then the Hamiltonian is rewritten as, 

\begin{eqnarray}
H  & = & (E_{1}+ \frac{E_{2}x^{2}}{1-x^{2}}-\frac{2E_{3}x\cos\phi}{\sqrt{1-x^{2}}})|w \rangle \langle w|  \nonumber \\ && + \frac{E_{2}}{1-x^{2}}|\psi \rangle\langle \psi |  \nonumber \\
& & + \frac{1}{\sqrt{1-x^{2}}}(E_{3}e^{i\varphi}-\frac{E_{2}x}{\sqrt{1-x^{2}}})|w\rangle \langle \psi|  \nonumber \\&&+\frac{1}{\sqrt{1-x^{2}}}(E_{3}e^{-i\varphi}-\frac{E_{2}x}{\sqrt{1-x^{2}}})|\psi \rangle \langle w|  \nonumber \\
\end{eqnarray} 

We require that this Hamiltonian should perform quantum search, then the Hamiltonian behaves expectedly,

\begin{eqnarray}
 H=E(|w \rangle \langle w| + |\psi \rangle \langle \psi|)  + \epsilon (e^{i\phi}|w \rangle \langle \psi| + e^{-i \phi} |\psi \rangle \langle w|) 
\end{eqnarray}

where $E_{1} = E(1+x^{2})+2\epsilon x \cos\phi$, $E_{2}=E(1-x^{2})$ and $E_{3}e^{\pm i\varphi} = \sqrt{1-x^{2}}(Ex +\epsilon e^{\pm i \phi})$. Thus we have shown that the quantum search Hamiltonian describing the generalized interaction between the initial state and the target state is the generalized quantum search Hamiltonian.  \\
 We now wish to discuss the source of the maximum speedup. First we compare oracular Hamiltonians in the Farhi and Gutmann Hamiltonian and the generalized quantum search Hamiltonian. In the former Hamiltonian, the oracular Hamiltonian is $H_{w} = E|w\rangle \langle w|$ and in the latter one, it is $H_{w} = (E+2\epsilon x \cos\phi)|w\rangle \langle w|$. That is, the difference between  magnitudes of two oracular Hamiltonians is little by the amount of $2\epsilon x \cos \phi$. Moreover, for the $O(1)$ time quantum search Hamiltonian, oracular Hamiltonians are the same. This implies that the oracle cannot be the source of the $O(1)$ time speedup. \\
 Therefore we notice that the factors, such as  the global interaction(or equivalently the mutual entanglement) between the target state and the initial state and the quantum coherence due to the phase alignment $\phi = n\pi$, are the sources of the maximum speedup. In the generalized quantum search Hamiltonian, the target state has much more interaction with the initial state compared to the interaction of the Farhi and Gutmann Hamiltonian. Evidently, the interaction $E_{3}e^{\pm i\varphi} = \sqrt{1-x^{2}}(Ex +\epsilon e^{\pm i \phi})$ is much larger than the interaction $Ex\sqrt{1-x^{2}}$. The difference between two interactions is about $\epsilon$. Then, as we have shown, the exponential speedup appears only with the condition of maximal interaction $\epsilon = E$. Also the other source of the maximum speedup is the quantum coherence due to the phase alignment $\phi = \pi/2$ if $\epsilon$ is suitably large. The type 4 Hamiltonian shows the power of the phase alignment. The $O(1)$ time speedup appears only when the conditions such as maximal interaction and phase alignment are satisfied. \\
 This implies that entanglement and quantum coherence are the sources of the speedup in a continuous time quantm search algorithm.

\section*{Acknowledgement}
J. Bae is supported in part by the Hanyang University Fellowship and Y. Kwon is supported in part by the Fund of Hanyang University.

\end{document}